\documentclass[conference,a4paper]{IEEEtran}

\usepackage[utf8]{inputenc}
\usepackage[T1]{fontenc}

\usepackage{amsmath,amssymb,amsfonts}
\usepackage{bm}
\usepackage{braket}
\usepackage{cite}
\usepackage{graphicx}
\usepackage{booktabs}
\usepackage{xcolor}
\usepackage{comment}

\DeclareMathOperator{\Tr}{Tr}

\title{Transient State Preservation of a Dissipative Qubit via Lyapunov Control}

\author{
\IEEEauthorblockN{Ilan Sabaj, Pablo Monz\'on}
\IEEEauthorblockA{
Department of Systems and Control, Universidad de la Rep\'ublica\\
\{isabaj,monzon\}@fing.edu.uy
}
}

\begin{document}

\maketitle

 \begin{abstract}
Preserving quantum states in the presence of environmental decoherence remains a fundamental challenge for quantum information processing. This work investigates whether Hamiltonian Lyapunov feedback can transiently preserve the state of an open qubit by reshaping its trajectory under Markovian dissipation. A state-error Lyapunov function is adopted to derive an analytical feedback law for Lindblad dynamics, ensuring a nonincreasing Lyapunov function whenever the control sensitivities are nonzero. The analysis identifies the geometric conditions under which the feedback becomes ineffective, showing that control authority vanishes when the current and target Bloch vectors are collinear. Numerical simulations are performed for amplitude damping and pure dephasing using selected target states. Performance is evaluated through fidelity, accumulated infidelity, dwell time above a fidelity threshold, coherence, and control effort. The results demonstrate that the proposed feedback cannot eliminate irreversible dissipation or achieve universal asymptotic stabilization, but it can provide transient benefits for selected target–channel combinations. In particular, under amplitude damping, the controller significantly reduces accumulated infidelity and substantially extends the interval during which the system remains above a prescribed fidelity threshold. In the pure dephasing cases examined, improvements in fidelity are generally limited, although transient coherence preservation is observed for the oblique target considered, revealing a trade-off between coherence and state fidelity. These findings indicate that Lyapunov-based Hamiltonian feedback is best suited as a transient preservation mechanism and motivate its application to gate operations and preservation of prepared states during finite time intervals before their use or readout.

 \end{abstract}

\section{Introduction}
\label{sec:introduction}

Quantum control uses external fields to steer quantum systems. For example, superconducting and spin qubits are commonly driven by microwave voltage pulses and oscillating magnetic or electric fields, respectively, while environmental couplings degrade the encoded information \cite{krantz_2019,loss_divincenzo_1998}. In quantum information processing, qubit states must be prepared, preserved, and transformed with high fidelity. We model environmental interactions by a Markovian Lindblad equation, whose Hamiltonian and dissipative terms describe unitary control and environmental effects, respectively \cite{breuer_petruccione_2002,gorini_kossakowski_sudarshan_1976,lindblad_1976}. Because the controls enter only through the Hamiltonian, they can reshape the trajectory but cannot directly cancel the dissipator. Lyapunov control techinques derive analytical state-feedback laws by choosing a target-error function $V(\rho)$ and designing controls that make it nonincreasing, avoiding repeated finite-horizon optimization. It has been widely studied for closed-system tasks such as state transfer, stabilization, mixed-state control, and quantum-operation implementation \cite{kuang_cong_2008,wang_schirmer_2009,yu_2024}. In open systems, prior work has often targeted decoherence-free subspaces, which are insensitive to selected environmental processes \cite{yi_2009,wang_wang_yi_2010}.

In this work, we investigate whether Hamiltonian feedback can delay dissipation-induced departure from a target by reshaping the qubit trajectory. We derive a Lyapunov feedback law and assess how its transient effectiveness depends on the dissipative process and target state. This is a theoretical and numerical study of an idealized effective two-level model with prescribed Markovian dissipation and assumed access to three Hamiltonian control directions. A platform-specific realization would require identifying the corresponding actuators, their amplitude and bandwidth constraints, and the validity of the two-level and dissipation approximations. Control effort is reported alongside state-preservation performance, but the feedback is designed to decrease the chosen Lyapunov function, without jointly optimizing control effort or other performance metrics. The remainder is organized as follows. Section~\ref{sec:preliminaries} introduces the model, Sections~\ref{sec:lyapunov_functions}--\ref{sec:feedback_law} derive the Lyapunov feedback law and characterize its limitations, and Section~\ref{sec:results} presents numerical examples for energy relaxation and pure dephasing.

\section{Preliminaries on controlled and open quantum systems}
\label{sec:preliminaries}

Throughout the paper, units with $\hbar=1$ are used. For a matrix $A$, $A^\dagger$ denotes its conjugate transpose, $\Tr(A)$ its trace, and $I$ the identity matrix of the appropriate dimension. The commutator and anticommutator are denoted by $[A,B]=AB-BA$ and $\{A,B\}=AB+BA$, respectively.

We consider the controlled Hamiltonian:
\begin{equation}
H_{\bm u}(t)
:=
H_0+\sum_k u_k(t)H_k.
\label{eq:controlled_hamiltonian}
\end{equation}
Here, $H_0$ is the drift Hamiltonian, $H_k$ are the available control Hamiltonians, and $u_k(t)\in\mathbb R$ are their amplitudes \cite{dalessandro_2021,kuang_cong_2008,cong_meng_2013}. A quantum state is described by a Hermitian positive semidefinite density operator $\rho$ with unit trace. Its diagonal and off-diagonal entries encode measurement probabilities and coherences. Pure states have $\rho=\ket{\psi}\bra{\psi}$ and purity $P(\rho):=\Tr(\rho^2)=1$, whereas mixed states have $P(\rho)<1$.

To include interactions with an external environment, we employ the Lindblad master equation, which generalizes the unitary dynamics to Markovian open-system evolution \cite{breuer_petruccione_2002,gorini_kossakowski_sudarshan_1976,lindblad_1976}:
\begin{equation}
\dot{\rho}
=
-i[H_{\bm u}(t),\rho]
+
\sum_j
\left(
L_j\rho L_j^\dagger
-
\frac{1}{2}\{L_j^\dagger L_j,\rho\}
\right).
\label{eq:lindblad}
\end{equation}
The operators $L_j$ define the prescribed dissipator $\mathcal{D}[rho]$. The deterministic Lindblad equation incorporates the statistical effects of relaxation and dephasing; dissipative rates are assumed known and constant, and model-parameter uncertainty is not considered.

The state of a single qubit admits a useful geometric representation through the Bloch vector $\bm r$, defined by
\begin{equation}
\rho
=
\frac{1}{2}\left(I+\bm r\cdot\bm\sigma\right),
\label{eq:bloch_representation}
\end{equation}
where $\bm r\in\mathbb R^3$, $\bm\sigma=(\sigma_x,\sigma_y,\sigma_z)$ is the vector of Pauli matrices, and $\bm r\cdot\bm\sigma=\sum_{a=x,y,z}r_a\sigma_a$. Physical states satisfy $|\bm r|\leq1$, with $P(\rho)=(1+|\bm r|^2)/2$; pure states have $|\bm r|=1$.

In the Bloch representation, Hamiltonian terms rotate $\bm r$ without changing its norm, whereas dissipation may change both its direction and norm.

We assume full state and model information. Given a known initial state and model, the feedback waveform can be computed offline and applied open loop \cite{grivopoulos_bamieh_2003}. Tomography on repeated preparations can support calibration and validation \cite{nielsen_chuang_2010}, whereas real-time feedback requires state estimation accounting for measurement backaction and latency.

\section{Lyapunov Functions and Derivative Structure}
\label{sec:lyapunov_functions}

Let $\rho_f$ denote the target state. We consider two natural Lyapunov candidate functions defined directly on density operators. One overlap-based function, and one state-error based function:
\begin{equation}
V_{\mathrm{ov}}(\rho)=1-\Tr(\rho\rho_f),\quad
V_{\mathrm{err}}(\rho)
=
\frac{1}{2}\Tr\left[(\rho-\rho_f)^2\right].
\label{eq:Verr_Vov_def}
\end{equation}

With the Hilbert--Schmidt inner product defined as $\langle A,B\rangle_{\mathrm{HS}}=\Tr(A^\dagger B)$, $V_{\mathrm{ov}}$ is related to the Hilbert--Schmidt inner product between the current and target states, whereas $V_{\mathrm{err}}=\frac{1}{2}\|\rho-\rho_f\|_{\mathrm{HS}}^2$ is one half of their squared Hilbert--Schmidt distance. In the Bloch representation introduced in Section~\ref{sec:preliminaries}, these quantities are expressed as
\begin{equation}
V_{\mathrm{ov}}(\rho)
=
\frac{1}{2}(1-\bm r\cdot\bm r_f),
\qquad
V_{\mathrm{err}}(\rho)
=
\frac{1}{4}|\bm r-\bm r_f|^2.
\end{equation}
$V_{\mathrm{ov}}$ is minimized at $\rho=\rho_f$ when $\rho_f$ is a pure state. For a mixed target, it is minimized at the eigenstate associated with the largest eigenvalue of $\rho_f$. In contrast, $V_{\mathrm{err}}$ is minimized at $\rho=\rho_f$ for any target state. Given a state $\rho$, both Lyapunov functions induce the same control direction. In the remainder of the paper, we use $V(\rho)=V_{\mathrm{err}}(\rho)$.

Since $\rho_f$ is constant, differentiating $V$ along the controlled Lindblad dynamics in \eqref{eq:lindblad} gives
\begin{equation}
\dot V(\rho)
=
\Tr\left[\dot\rho(\rho-\rho_f)\right].
\end{equation}
Substituting $\dot\rho$ from \eqref{eq:lindblad} into this expression and separating the uncontrolled and control-dependent contributions, the Lyapunov derivative can be written as
\begin{equation}
\dot V(\rho)
=
A(\rho)+\sum_k u_k(\rho)f_k(\rho),
\label{eq:general_lyapunov_derivative}
\end{equation}where
\begin{equation}
\begin{aligned}
A(\rho)
&=
\Tr\left(
\left[-i[H_0,\rho]+\mathcal{D}[\rho]\right]
(\rho-\rho_f)
\right),\\
f_k(\rho)
&=
-i\Tr\left(
[H_k,\rho](\rho-\rho_f)
\right).
\end{aligned}
\label{eq:A_fk_def}
\end{equation}
The term $A(\rho)$ collects the effects of the drift Hamiltonian and the dissipator, whereas $f_k(\rho)$ describes the instantaneous effect of the $k$th Hamiltonian control on the selected Lyapunov function.

\section{Feedback Law}
\label{sec:feedback_law}
A standard Lyapunov-control design is to choose each control amplitude proportional to the negative of its associated control sensitivity $f_k(\rho)$ \cite{kuang_cong_2008,wang_schirmer_2009}. With this choice, the control-dependent contribution $\sum_k u_k(\rho)f_k(\rho)$ produces a negative term proportional to $\sum_k f_k^2(\rho)$. 

The proportionality factor is selected so as to cancel the uncontrolled contribution $A(\rho)$. Under the full-state and model-information assumptions stated in Section~\ref{sec:preliminaries}, the quantities $A(\rho)$ and $f_k(\rho)$ can be evaluated at each instant. Thus, whenever
$\sum_j f_j^2(\rho)>0$, we choose
\begin{equation}
u_k(\rho)
=
-\left(
\frac{A(\rho)}
{\sum_j f_j^2(\rho)}
+
\kappa
\right)f_k(\rho),
\qquad
\kappa>0,
\label{eq:control_law}
\end{equation}
where $\kappa$ is a feedback gain. Substituting \eqref{eq:control_law} into the Lyapunov derivative gives
\begin{align}
\dot V(\rho)
&=
A(\rho)
+
\sum_k u_k(\rho)f_k(\rho)
=
-\kappa\sum_k f_k^2(\rho)
\leq 0.
\label{eq:dotV_negative}
\end{align}
Therefore, wherever the feedback law is well defined and at least one control sensitivity is nonzero, the state-error function decreases strictly. The feedback has no instantaneous effect on $V$ when all the control sensitivities vanish. To characterize these states, cyclicity of the trace gives $f_k(\rho) = i\Tr(H_k[\rho,\rho_f])$. If we assume that the available controls can generate independent Hamiltonian directions along the three Pauli axes---which in the Bloch-vector representation is equivalent to assuming that they can generate rotations of $\bm r$ about the $x$, $y$, and $z$ axes---then
\begin{equation}
f_k(\rho)=0,\ \forall k
\quad\Longleftrightarrow\quad
[\rho,\rho_f]=0.
\label{eq:fk_zero_commutation}
\end{equation}
Formally, this assumption requires that the traceless parts of the available control Hamiltonians span the real vector space of traceless Hermitian $2\times2$ matrices. In the Bloch representation introduced in Section~\ref{sec:preliminaries},
\begin{equation}
[\rho,\rho_f]
=
\frac{i}{2}
(\bm r\times\bm r_f)\cdot\bm\sigma.
\end{equation}
Therefore, for a nonzero target Bloch vector $\bm r_f$, the feedback sensitivities vanish simultaneously exactly when $\bm r$ and $\bm r_f$ are collinear. This behavior is geometrically natural: Hamiltonian controls induce rotations of $\bm r$ without changing its norm, and when $\bm r$ and $\bm r_f$ are collinear, no infinitesimal rotation can decrease the distance to the target to first order.

The feedback expression in \eqref{eq:control_law} is singular when all $f_k(\rho)$ vanish and may produce excessively large amplitudes near this set. In the numerical implementation, controls are switched off whenever $\sum_k f_k^2(\rho)$ falls below a prescribed threshold. The switching rule and threshold value are specified in Section~\ref{sec:numerical_implementation}. When the controls are switched off, the Lyapunov derivative reduces to $\dot V(\rho)=A(\rho)$, whose sign is not constrained by the feedback design. Accordingly, the strict decrease established above applies only while the feedback law is active.

The proposed feedback law should therefore be interpreted as a transient control mechanism: whenever the available Hamiltonian directions have a nonzero instantaneous effect on $V$, they are used to reduce the state error. Its effectiveness is limited near states at which these control sensitivities vanish and depends on the interplay between the target state, the available control Hamiltonians, and the dissipative dynamics. Section~\ref{sec:results} investigates this behavior for different target states and dissipative processes.

\section{Numerical Implementation}
\label{sec:numerical_implementation}

We take the available Hamiltonian control directions as the three Pauli axes:
\begin{equation}
 H_x=\frac{\sigma_x}{2},\qquad H_y=\frac{\sigma_y}{2},\qquad H_z=\frac{\sigma_z}{2}.
 \label{eq:numerical_hamiltonians}
\end{equation}
These directions span every traceless Hermitian qubit Hamiltonian, whereas an identity component has no dynamical effect. We set $H_0=0$ in the simulations. A known constant traceless drift could be compensated by control offsets provided sufficient amplitude headroom is available; such offsets would consume part of the available control budget.

We consider amplitude damping (AD), which models spontaneous energy relaxation toward the lower-energy state $\ket{0}$, and phase damping (PD), which degrades the off-diagonal coherences of $\rho$ without energy exchange. The corresponding collapse operators are
\begin{equation}
\begin{aligned}
 L_{\mathrm{AD}}&=\sqrt{\gamma_{\mathrm{AD}}}\ket{0}\bra{1}=\sqrt{\gamma_{\mathrm{AD}}}\begin{bmatrix}0&1\\0&0\end{bmatrix},\\
 L_{\mathrm{PD}}&=\sqrt{\frac{\gamma_{\mathrm{PD}}}{2}}\,\sigma_z=\sqrt{\frac{\gamma_{\mathrm{PD}}}{2}}\begin{bmatrix}1&0\\0&-1\end{bmatrix}.
\end{aligned}
\label{eq:numerical_collapse_operators}
\end{equation}
With this normalization, $\gamma_{\mathrm{PD}}$ is the coherence-decay rate. In the absence of Hamiltonian terms, AD drives every state toward the $\ket{0}$ north pole state of the Bloch sphere, while PD projects every Bloch vector onto its $z$ axis component. 

Each dissipative process is characterized by a rate $\gamma$ with units of $\mathrm{s}^{-1}$. We introduce the dimensionless time $\tau=\gamma_{\mathrm{ref}}t$, where $\gamma_{\mathrm{ref}}$ is the rate of the active process, so simulation time is measured in units of $1/\gamma_{\mathrm{ref}}$. All reported single-channel cases therefore use a unit dimensionless dissipation rate.

The feedback is implemented as sampled-data, piecewise-constant control. At time $\tau_n$, the sensitivities and ideal amplitudes in \eqref{eq:control_law} are evaluated from $\rho(\tau_n)$. To avoid division by a nearly vanishing sensitivity, all controls are set to zero when $S(\rho):=\sum_k f_k^2(\rho)\leq\varepsilon_s$. Control amplitudes are clipped to $[-u_{\max},u_{\max}]$ to maintain reasonable control effort, and the resulting Hamiltonian is held fixed on $[\tau_n,\tau_{n+1})$. QuTiP's \texttt{mesolve} solver integrates the Lindblad equation with adaptive internal steps, whereas $\Delta\tau_{\mathrm{ctrl}}=\tau_{n+1}-\tau_n$ is only the feedback update period. Because the applied fields are sampled and clipped, they need not satisfy the ideal identity $\dot V=-\kappa S$ continuously. The nominal parameters are $\kappa=0.5$, $u_{\max}=2$, $\Delta\tau_{\mathrm{ctrl}}=0.01$, $\varepsilon_s=10^{-10}$, $\gamma_{\mathrm{AD}}=\gamma_{\mathrm{PD}}=1$, and final time $T=8$. They were selected through one-factor-at-a-time sweeps over selected AD and PD cases with different target states. Increasing $u_{\max}$ beyond $2$ yielded only modest further improvement but substantially increased the control effort $\mathcal E_u =\int_0^T\sum_k u_k^2d\tau$. Results were nearly unchanged for $\Delta\tau_{\mathrm{ctrl}}\in[0.005,0.1]$. We choose $\varepsilon_s=10^{-10}$ as the largest tested value that preserves the converged behavior.

Unit-trace preservation, Hermiticity, and positive semidefiniteness were checked along every nominal trajectory. The maximum numerical deviations were of order $10^{-16}$, consistent with floating-point limits.

\section{Results}
\label{sec:results}

We quantify target proximity by the squared Uhlmann fidelity \cite{nielsen_chuang_2010},
\begin{equation}
 F(\rho,\rho_f)=\left[\Tr\sqrt{\sqrt{\rho_f}\rho\sqrt{\rho_f}}\right]^2.
 \label{eq:uhlmann_fidelity}
\end{equation}
For the pure targets considered here, $F=\Tr(\rho\rho_f)=(1+\bm r\cdot\bm r_f)/2$. Transient performance is measured by accumulated infidelity and dwell time above a fidelity threshold,
\begin{equation}
 \mathcal J_F=\int_0^T[1-F(\tau)]\,d\tau,\qquad t_{F\geq F_*}=\int_0^T\mathbf{1}_{\{F(\tau)\geq F_*\}}\,d\tau.
 \label{eq:transient_metrics}
\end{equation}
We use $F_*=0.75$ and also report final fidelity, purity, computational-basis coherence $C_{l_1}=2|\rho_{01}|$, and control effort $\mathcal E_u=\int_0^T\sum_k u_k^2d\tau$.

The \emph{exact} cases start at $\rho(0)=\rho_f$. If dissipation preserves collinearity with the target, all $f_k$ vanish and the feedback cannot start; we then also test a $5^\circ$ polar and azimuthal perturbation of the initial Bloch vector (\emph{perturbed}). Positive $\Delta\mathcal J_F$ in Table~\ref{tab:numerical_summary} indicates improvement over the matched uncontrolled trajectory.

The selected targets illustrate distinct geometric mechanisms: persistent collinearity for exact AD--$\ket{1}$ and PD--$\ket{+}$, dissipation-induced departure from collinearity for AD--$\ket{+}$, and feedback activation for an oblique PD target. These cases probe how target geometry affects control activation and transient performance

\begin{figure}[!t]
    \centering
    \includegraphics[width=0.84\columnwidth,trim=0 0 0 38bp,clip]{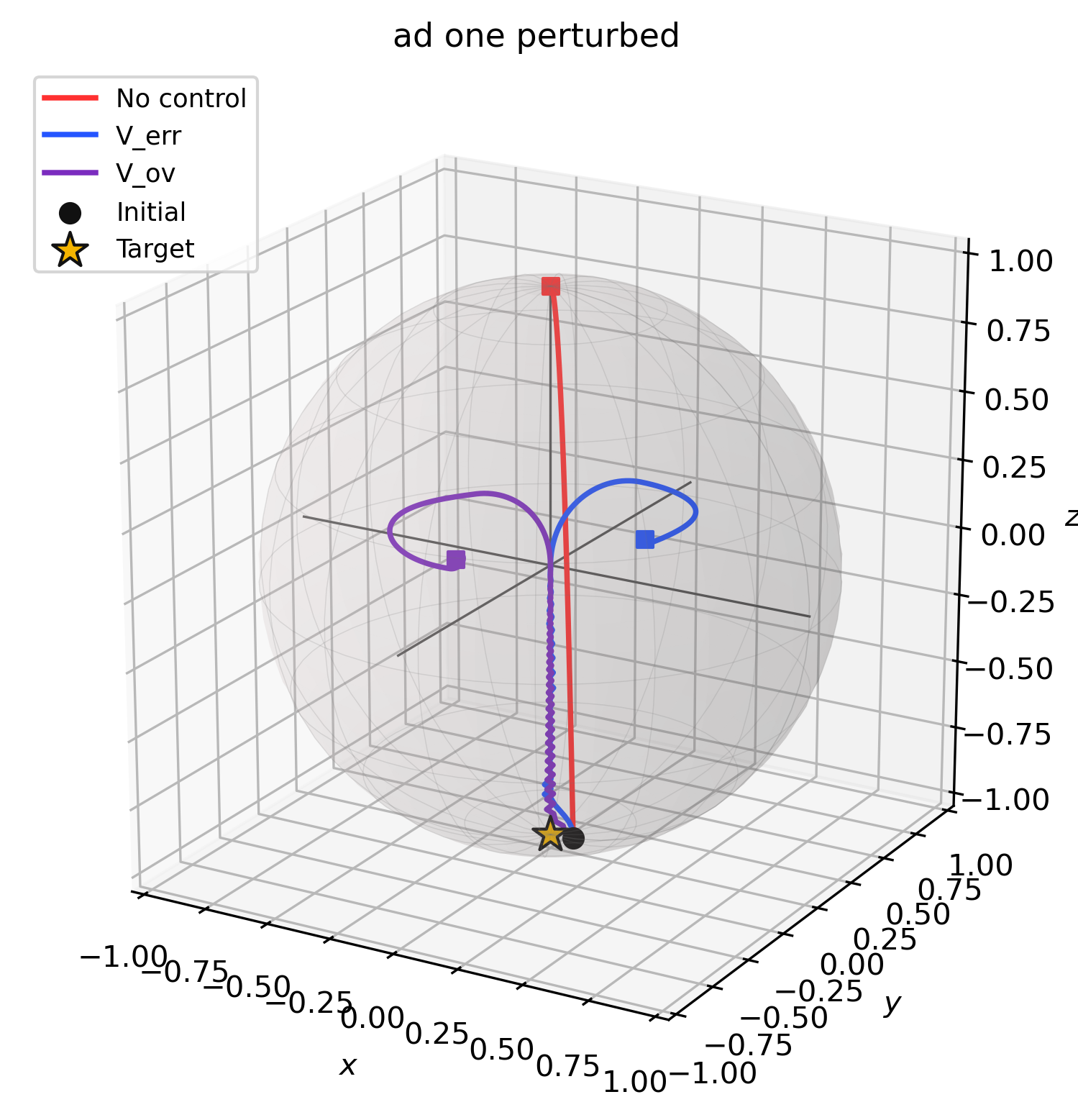}
    \par\vspace{-0.7em}{\small (a)}\par\vspace{0.3em}
    \includegraphics[width=0.8\columnwidth]{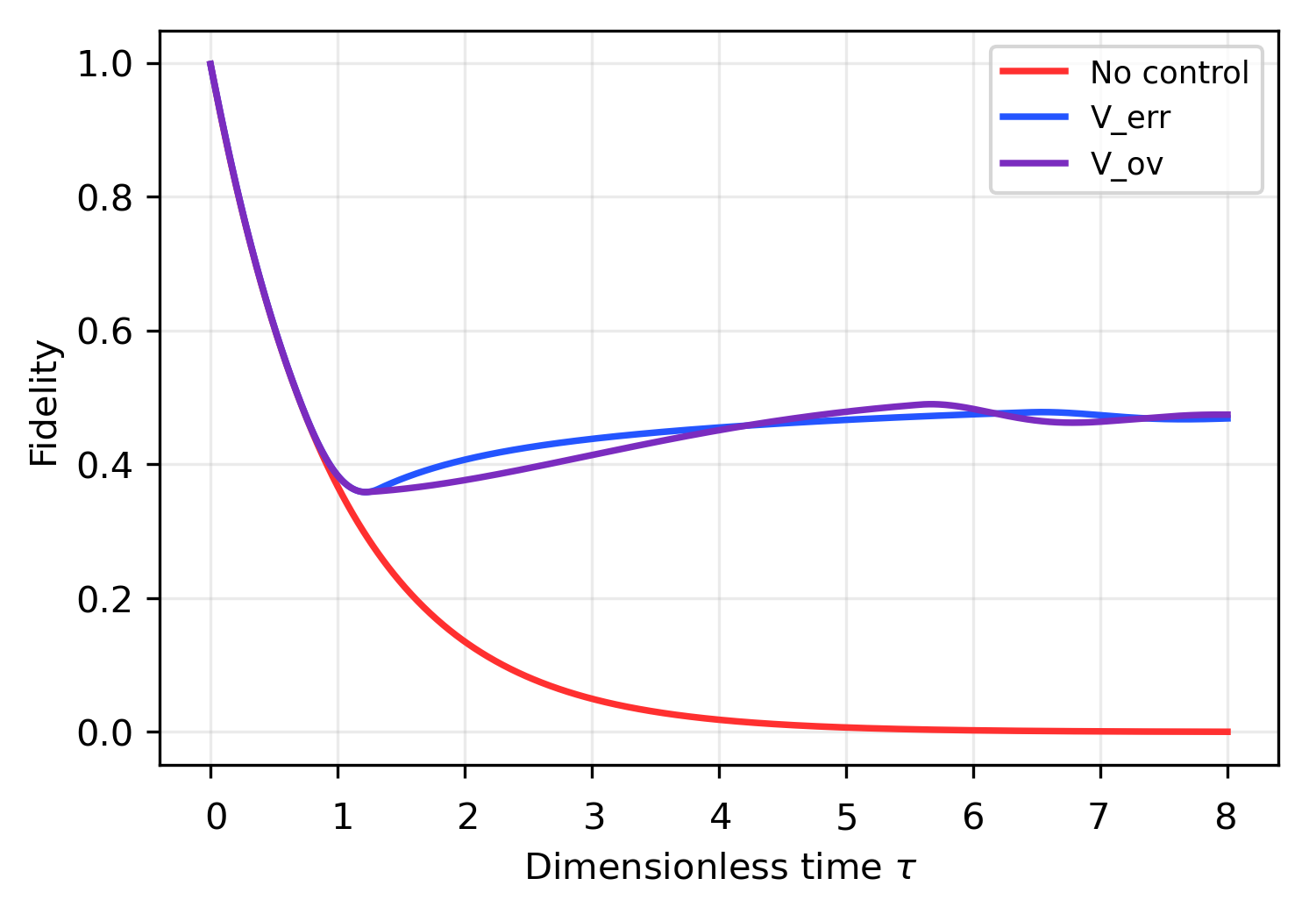}
    \par\vspace{-0.7em}{\small (b)}\par\vspace{0.3em}
    \includegraphics[width=0.8\columnwidth]{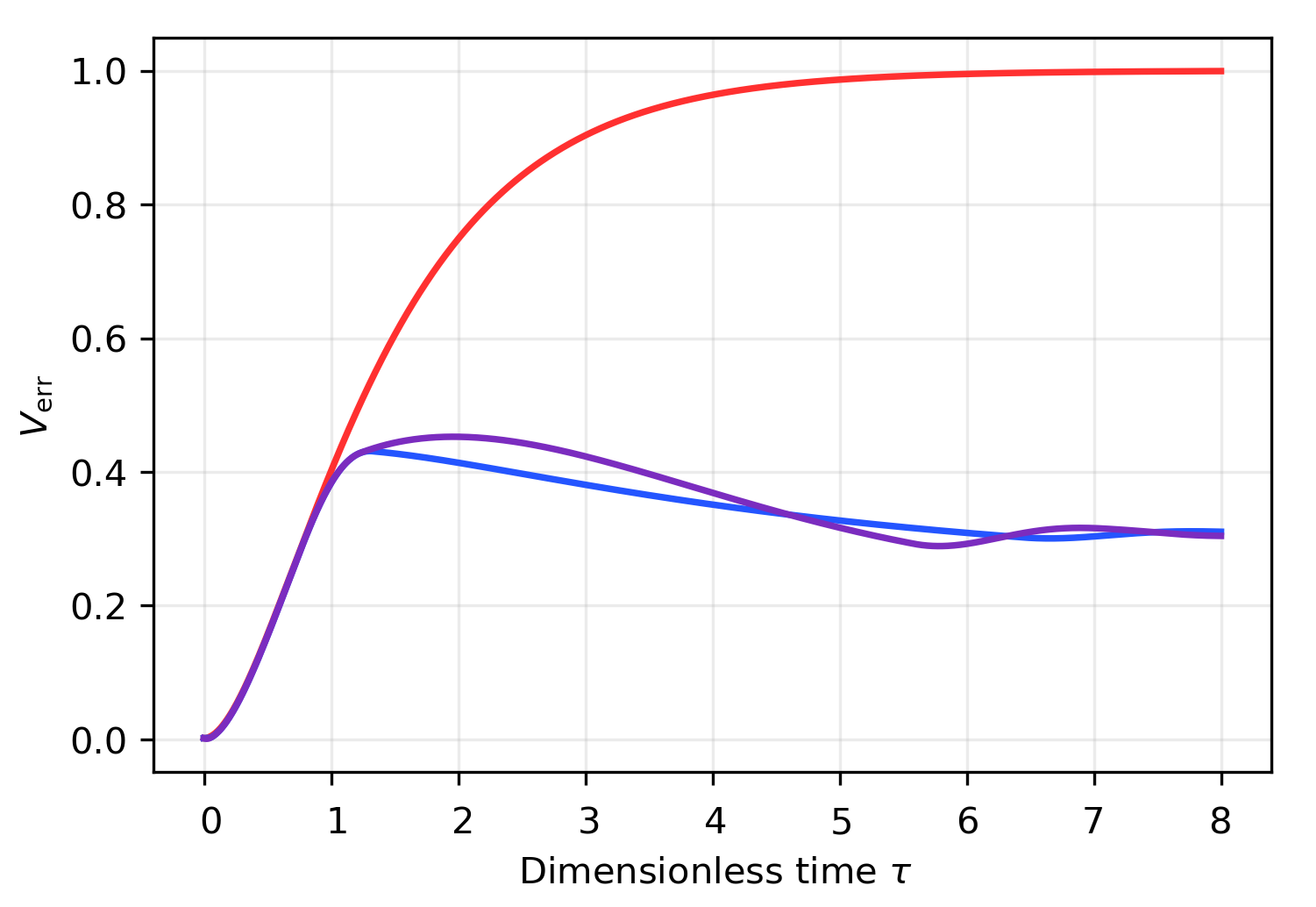}
    \par\vspace{-0.7em}{\small (c)}
    \caption{AD with a perturbed $\ket{1}$ target: (a) Bloch trajectories, (b) fidelity, and (c) state error. Feedback reduces accumulated infidelity but barely changes the moderate-fidelity interval.}
    \label{fig:ad_one_perturbed}
\end{figure}

In the AD--$\ket{1}$ case, the target is opposite to the attractor $\ket{0}$. From the exact target the state remains on the $z$ axis and the feedback is inactive. The perturbation used in Fig.~\ref{fig:ad_one_perturbed} breaks this symmetry: both $V_{\mathrm{err}}$- and $V_{\mathrm{ov}}$-based laws reduce $\mathcal J_F$ by about $39\%$, although neither materially extends $t_{F\geq0.75}$. Their sensitivity vectors coincide, but different uncontrolled terms and normalizations produce distinct Bloch trajectories with similar fidelity and error curves.

\begin{figure}[!t]
    \centering
    \includegraphics[width=0.84\columnwidth,trim=0 0 0 38bp,clip]{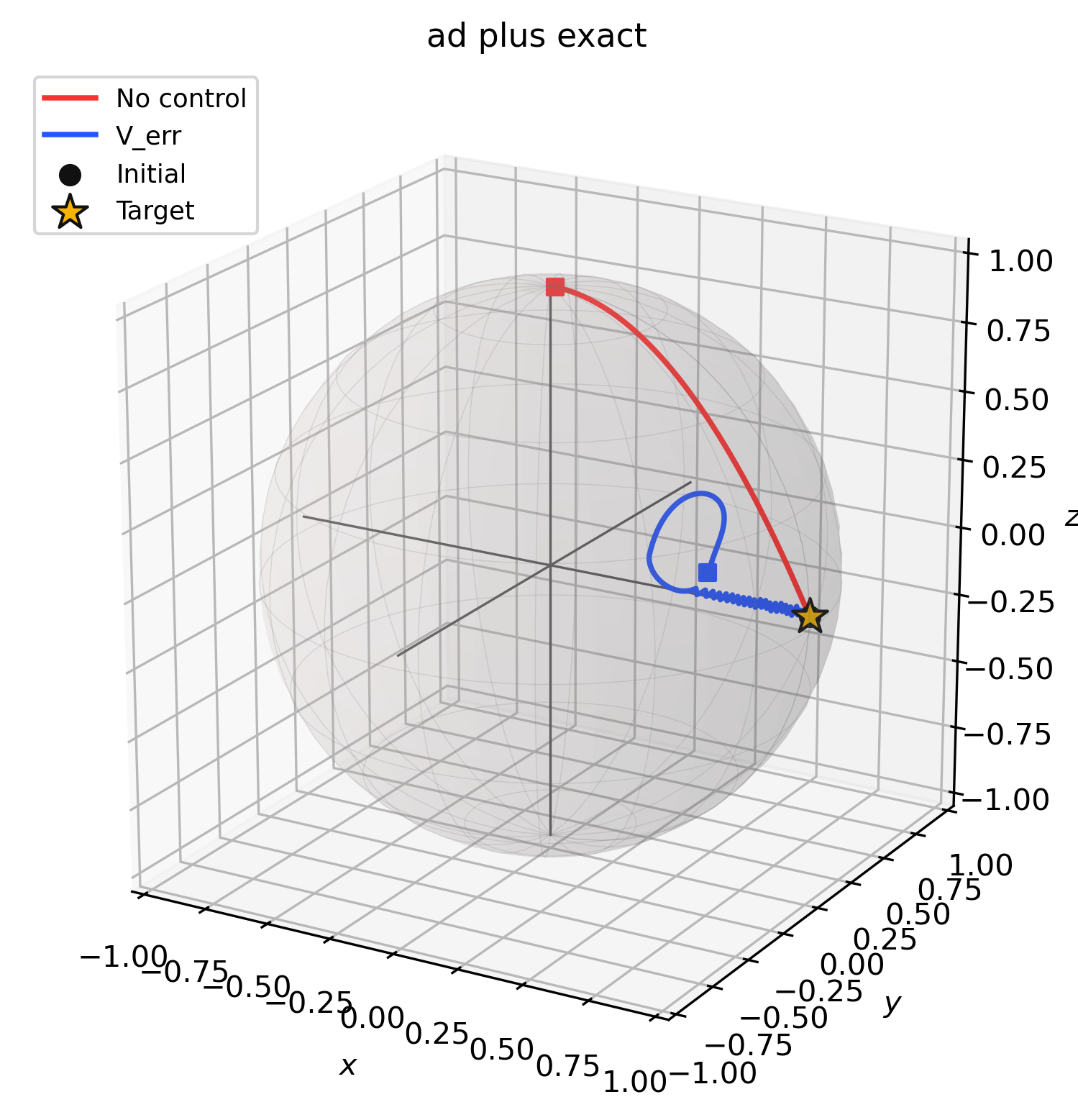}
    \par\vspace{-0.7em}{\small (a)}\par\vspace{0.3em}
    \includegraphics[width=0.8\columnwidth]{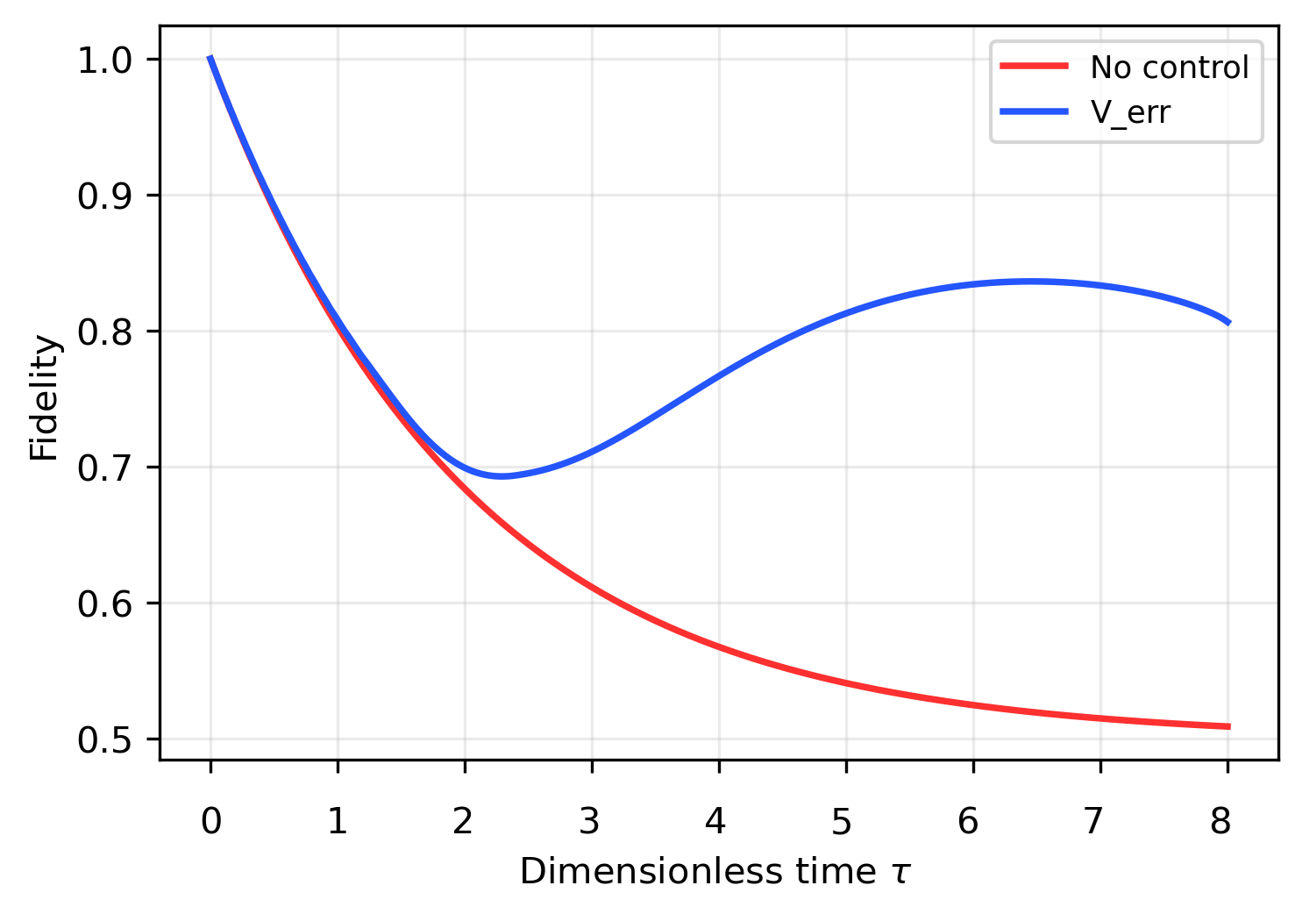}
    \par\vspace{-0.7em}{\small (b)}\par\vspace{0.3em}
    \caption{AD with the exact $\ket{+}$ target: (a) Bloch trajectories and (b) fidelity. Feedback substantially extends the interval above $F=0.75$.}
    \label{fig:ad_plus_exact}
\end{figure}

For exact AD--$\ket{+}$, AD immediately creates a $z$ component and activates the controller after the first update. As Fig.~\ref{fig:ad_plus_exact} shows, mean fidelity increases from $0.623$ to $0.794$, final fidelity from $0.509$ to $0.807$, and $t_{F\geq0.75}$ from $1.39$ to $5.73$.

For PD--$\ket{+}$, the exact trajectory remains on the target axis and the controller is inactive. A $5^\circ$ perturbation activates it, but Hamiltonian rotations cannot restore the transverse radius removed by PD; $\mathcal J_F$ improves by only $0.1\%$, with negligible changes in final fidelity and dwell time.

\begin{figure*}[!t]
    \centering
    \begin{minipage}[c]{0.46\textwidth}
        \centering
        \includegraphics[width=0.92\linewidth,trim=0 0 0 38bp,clip]{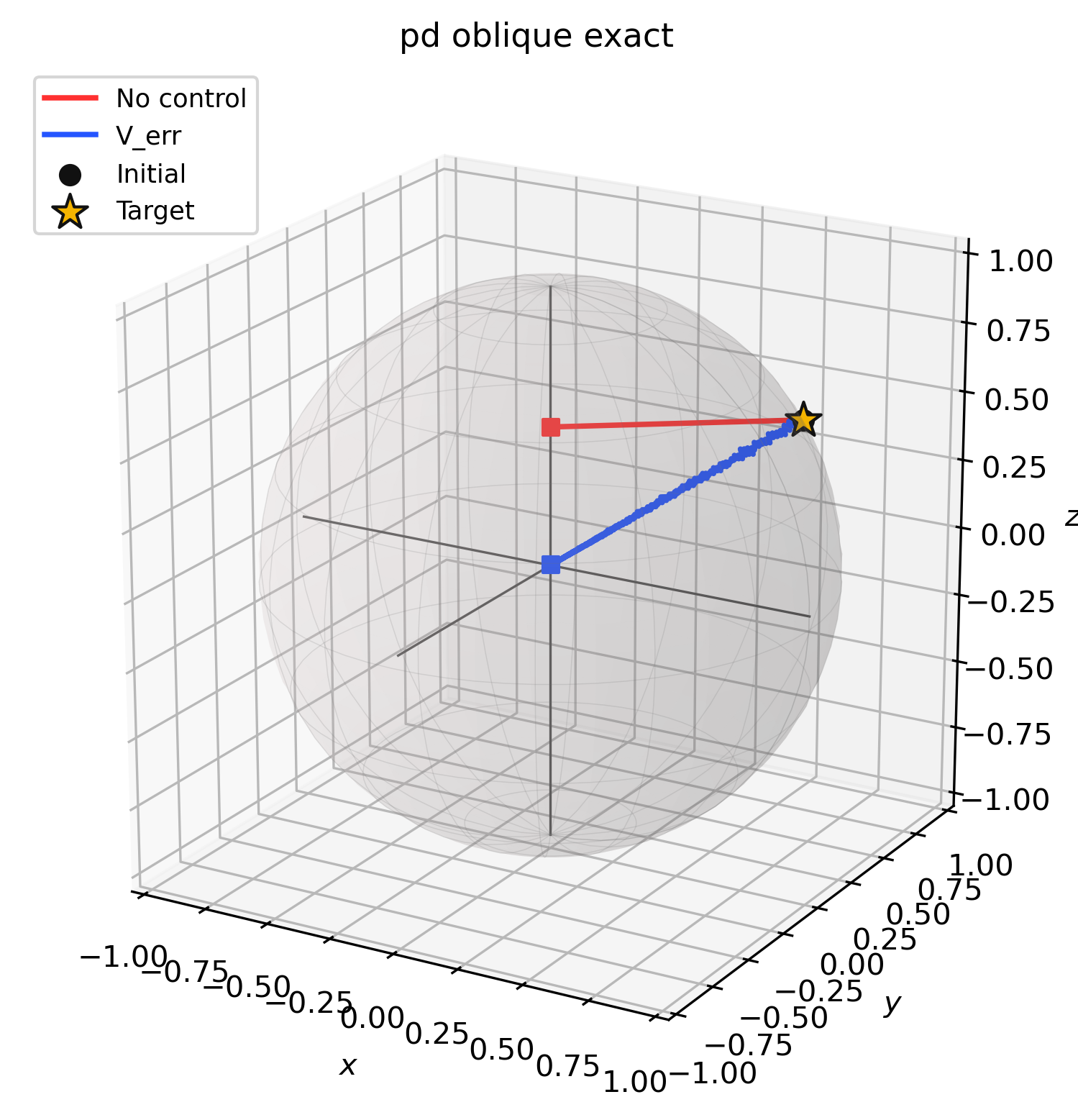}
        \par\vspace{-0.6em}{\small (a)}
    \end{minipage}\hfill
    \begin{minipage}[c]{0.49\textwidth}
        \centering
        \includegraphics[width=0.8\linewidth]{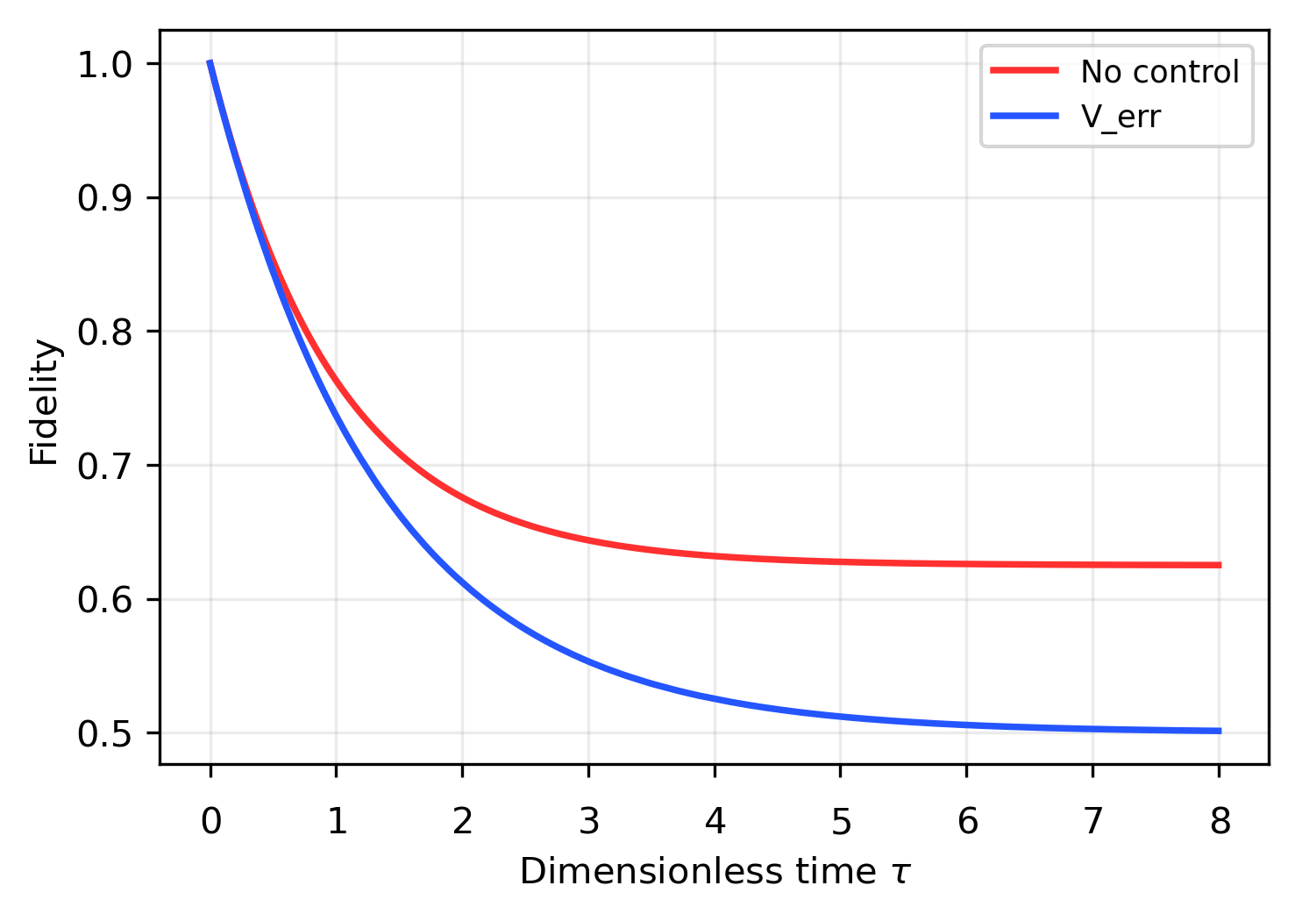}
        \par\vspace{-0.7em}{\small (b)}\par\vspace{0.1em}
        \includegraphics[width=0.8\linewidth]{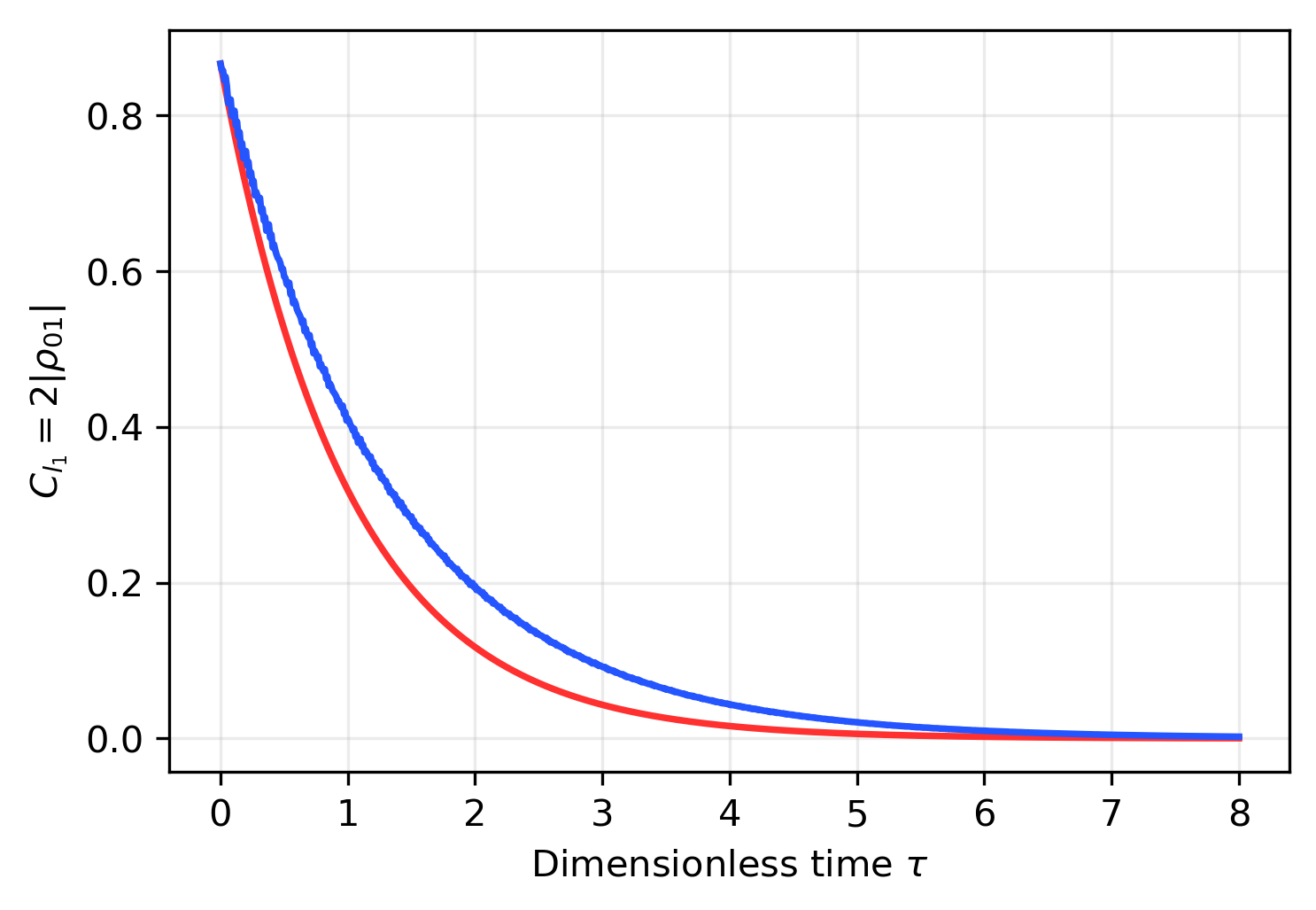}
        \par\vspace{-0.7em}{\small (c)}
    \end{minipage}
    \caption{PD with the exact oblique target $\ket{\psi_N}$: (a) Bloch trajectories, (b) fidelity, and (c) computational-basis coherence.}
    \label{fig:pd_oblique_exact}
\end{figure*}

For the oblique target
\begin{equation}
\begin{aligned}
 \ket{\psi_N}&=\cos\!\left(\frac{\theta}{2}\right)\ket{0}
 +e^{i\phi}\sin\!\left(\frac{\theta}{2}\right)\ket{1},\\
 (\theta,\phi)&=(60^\circ,35^\circ),
\end{aligned}
\label{eq:oblique_target}
\end{equation}
PD changes the Bloch-vector direction and activates the controller, producing the trade-off in Fig.~\ref{fig:pd_oblique_exact}. Control increases $\mathcal J_F$ by $26.9\%$ and drives the final purity toward $1/2$, but $C_{l_1}$ exceeds the uncontrolled value for almost the entire run, with maximum and integrated advantages of $0.096$ and $0.288$, respectively. Feedback repeatedly rotates the $z$ component preserved by PD toward the transverse plane, where it is damped; this transiently preserves coherence at the cost of Bloch-vector length and fidelity.

\begin{table*}[t]
\centering
\caption{Selected state-holding results over $T=8$.}
\label{tab:numerical_summary}
\resizebox{\textwidth}{!}{%
\begin{tabular}{llllrrrr}
\toprule
Channel & Target and initial state & Law & Behavior & $\Delta\mathcal J_F$ [\%] & $\Delta t_{F\geq0.75}$ & $F(T)$ & $\mathcal E_u$ \\
\midrule
AD & $\ket{1}$, perturbed & $V_{\mathrm{err}}$ & transient improvement & $39.5$ & $0.00$ & $0.469$ & $46.5$ \\
AD & $\ket{1}$, perturbed & $V_{\mathrm{ov}}$ & transient improvement & $38.8$ & $0.00$ & $0.474$ & $45.9$ \\
AD & $\ket{+}$, exact & $V_{\mathrm{err}}$ & delayed start, improvement & $45.5$ & $4.34$ & $0.807$ & $18.7$ \\
PD & $\ket{+}$, perturbed & $V_{\mathrm{err}}$ & negligible improvement & $0.1$ & $0.01$ & $0.500$ & $51.4$ \\
PD & $\ket{\psi_N}$, exact & $V_{\mathrm{err}}$ & fidelity--coherence trade-off & $-26.9$ & $-0.17$ & $0.501$ & $92.9$ \\
\bottomrule
\end{tabular}}
\end{table*}

\section{Conclusions}
\label{sec:conclusions}

Purely Hamiltonian feedback cannot undo irreversible dissipation or provide universal asymptotic state stabilization. However, results show that it can produce a moderately useful transient effect for selected target--channel pairs: it reduces accumulated infidelity in several AD cases and substantially extends the moderate-fidelity operating window for AD--$\ket{+}$. These transient benefits motivate further study of Lyapunov-based state transfer and of correction controls superimposed on nominal gate-driving fields. The geometric characterization and numerical benefits established here apply to a single qubit with prescribed dissipation and do not automatically extend to coupled multiqubit systems.
\bibliographystyle{IEEEtran} 

\begin{thebibliography}{99}

\bibitem{krantz_2019}
P. Krantz, M. Kjaergaard, F. Yan, T. P. Orlando, S. Gustavsson, and W. D. Oliver, ``A quantum engineer's guide to superconducting qubits,'' \emph{Appl. Phys. Rev.}, vol. 6, no. 2, Art. no. 021318, 2019.

\bibitem{loss_divincenzo_1998}
D. Loss and D. P. DiVincenzo, ``Quantum computation with quantum dots,'' \emph{Phys. Rev. A}, vol. 57, no. 1, pp. 120--126, 1998.

\bibitem{breuer_petruccione_2002}
H.-P. Breuer and F. Petruccione, \emph{The Theory of Open Quantum Systems}. Oxford, U.K.: Oxford Univ. Press, 2002.

\bibitem{gorini_kossakowski_sudarshan_1976}
V. Gorini, A. Kossakowski, and E. C. G. Sudarshan, ``Completely positive dynamical semigroups of $N$-level systems,'' \emph{J. Math. Phys.}, vol. 17, no. 5, pp. 821--825, 1976.

\bibitem{lindblad_1976}
G. Lindblad, ``On the generators of quantum dynamical semigroups,'' \emph{Commun. Math. Phys.}, vol. 48, no. 2, pp. 119--130, 1976.

\bibitem{kuang_cong_2008}
S. Kuang and S. Cong, ``Lyapunov control methods of closed quantum systems,'' \emph{Automatica}, vol. 44, no. 1, pp. 98--108, 2008.

\bibitem{wang_schirmer_2009}
X. Wang and S. G. Schirmer, ``Analysis of Lyapunov method for control of quantum states,'' \emph{IEEE Trans. Autom. Control}, vol. 55, no. 10, pp. 2259--2270, 2010.

\bibitem{yu_2024}
G.-H. Yu and H.-L. Yang, ``Quantum control based on three forms of Lyapunov functions,'' \emph{Chin. Phys. B}, vol. 33, no. 4, 2024.

\bibitem{yi_2009}
X. X. Yi, X. L. Huang, C. Wu, and C. H. Oh, ``Driving quantum systems into decoherence-free subspaces by Lyapunov control,'' \emph{Phys. Rev. A}, vol. 80, no. 5, Art. no. 052316, 2009.

\bibitem{wang_wang_yi_2010}
W. Wang, L. C. Wang, and X. X. Yi, ``Lyapunov control on quantum open systems in decoherence-free subspaces,'' \emph{Phys. Rev. A}, vol. 82, no. 3, Art. no. 034308, 2010.

\bibitem{nielsen_chuang_2010}
M. A. Nielsen and I. L. Chuang, \emph{Quantum Computation and Quantum Information}, 10th anniversary ed. Cambridge, U.K.: Cambridge Univ. Press, 2010.

\bibitem{dalessandro_2021}
D. D'Alessandro, \emph{Introduction to Quantum Control and Dynamics}, 2nd ed. Boca Raton, FL, USA: CRC Press, 2021.

\bibitem{cong_meng_2013}
S. Cong and F. Meng, ``A survey of quantum Lyapunov control methods,'' \emph{Sci. World J.}, vol. 2013, Art. no. 967529, 2013.

\bibitem{grivopoulos_bamieh_2003}
S. Grivopoulos and B. Bamieh, ``Lyapunov-based control of quantum systems,'' in \emph{Proc. 42nd IEEE Conf. Decision Control}, vol. 1, 2003, pp. 434--438.

\end{thebibliography}

\end{document}